\documentclass[mathpazo]{cicp}

\usepackage{lineno,hyperref}
\modulolinenumbers[5]

\usepackage{todonotes}
\usepackage{makeidx}
\usepackage{graphics}
\usepackage{color}
\usepackage{listings}
\usepackage{bm}
\usepackage{url}
\usepackage{booktabs}
\usepackage{amsmath} 
\usepackage{paralist}
\usepackage{amssymb}
\usepackage{amsthm}
\usepackage{subcaption} 
\usepackage{algorithm}
\usepackage{algorithmic}
\usepackage{lipsum}
\usepackage{upgreek}
\usepackage{hyperref}
\usepackage{graphicx}
\usepackage{dsfont}
  
\usepackage[percent]{overpic}

\newcommand*{\dd}{\partial}
\newcommand*{\ie}{\textit{i.e. }}
\newcommand*{\eq}{{\mbox{\tiny eq}}}

\newcommand*{\kb}{k_{\rm B}}

\newcommand*{\bld}{\boldsymbol}
\newcommand*{\diff}{ d }

\DeclareMathOperator{\arctanh}{arctanh}

\definecolor{amethyst}{rgb}{0.6, 0.4, 0.8}

\newtheorem*{theorem*}{Theorem:}

\begin{document} 

\title{Bjorken flow revisited: analytic and numerical solutions in flat space-time coordinates}


\author[D. Simeoni et.~al.]{
        Daniele Simeoni \affil{1,2,3}\comma\corrauth,
        Alessandro Gabbana \affil{4}, 
    and Sauro Succi\affil{5,6}}
\address{
\affilnum{1}\ Universit\`a di Ferrara and INFN-Ferrara, I-44122 Ferrara,~Italy                      \\
\affilnum{2}\ Bergische Universit\"at Wuppertal, D-42119 Wuppertal,~Germany                         \\ 
\affilnum{3}\ University of Cyprus, Physics department, CY-1678 Nicosia,~Cyprus                     \\
\affilnum{4}\ Eindhoven University of Technology, 5600 MB Eindhoven,~ The Netherlands               \\
\affilnum{5}\ Center for Life Nano Science @ La Sapienza, Italian Institute of Technology, Viale Regina Elena 295, I-00161 Roma,~Italy                                                                                         \\ 
\affilnum{6}\ Istituto Applicazioni del Calcolo, National Research Council of Italy, Via dei Taurini 19, I-00185 Roma,~Italy}

\emails{{\tt d.simeoni@stimulate-ejd.eu} (D.~Simeoni)}
 
 
\begin{abstract} 
  In this work we provide analytic and numerical solutions for the Bjorken flow, 
  a standard benchmark in relativistic hydrodynamics providing a simple model 
  for the bulk evolution of matter created in collisions between heavy nuclei.

  We consider relativistic gases of both massive and massless particles,
  working in a $(2+1)$ and $(3+1)$ Minkowski space-time coordinate system.
  The numerical results from a recently developed lattice kinetic scheme
  show excellent agreement with the analytic solutions.
\end{abstract}




\maketitle



\section{Introduction}\label{sec:1-intro}

In recent years, experimental data 
from the Relativistic Heavy-Ion Collider (RHIC) and the Large Hadron Collider (LHC)~\cite{
star-npa-2005, phobos-npa-2005, phenix-npa-2005, brahams-npa-2005, pasechnik-u-2017} 
has provided the first clear observation of the Quark-Gluon Plasma (QGP),
a deconfined phase of matter where quarks and gluons are effectively free beyond the 
nucleonic volume~\cite{satz-annurev-1985, karsch-npa-2002, busza-annurev-2018}.

Remarkably, the earliest stages following the heavy-ion collisions present collective
behaviors that can be described by the laws of fluid dynamics~\cite{jaiswal-ahep-2016}, 
and indeed these results have significantly boosted the interest in the study of viscous 
relativistic fluid dynamics~\cite{romatschke-book-2019}, both at the level 
of theoretical formulations as well as in the development of reliable numerical simulation methods.

Most of the numerical methods are based on Israel-Stewart theory~\cite{israel-prsl-1979}
or more recent second-order causal formalism~\cite{baier-jhep-2008}, with a few
example represented by MUSIC~\cite{schenke-prc-2010}, vSHASTA~\cite{molnar-epjc-2010}, 
ECHO-QGP~\cite{delzanna-epjc-2013}, and several more 
(see e.g.~\cite{karpenko-cpc-2014,okamoto-epj-2016} and citations therein).
Besides, mesoscopic approaches are often employed in order to study 
relativistic hydrodynamic systems, including lattice kinetic schemes~\cite{gabbana-pr-2020},
as well as Monte Carlo based methods~\cite{xu-prc-2007,plumari-prc-2012}.

The development of new numerical tools for the study of relativistic fluids 
is an active field of research. However evaluating and comparing the 
accuracy, stability and performance of the available solvers represent 
a challenging task, due to the fact that numerical benchmarks with an
analytic solution are rare and available only in very idealized situations.

Two of the most commonly used benchmarks are the Riemann problem~\cite{thompson-jofm-1986,marti-jofm-1994},
for which an analytic solution is available only in the inviscid limit
(see~\cite{bouras-prl-2009,bouras-npa-2009,bouras-prc-2010,bouras-jop-2010,gabbana-prc-2020}
for numerical results in the viscous regime), and the Bjorken flow~\cite{bjorken-prd-1983}
(as well as its generalization given by the Gubser flow~\cite{gubser-prd-2010}).

The Bjorken flow represents the simplest numerical setup in the QGP context.
Under the assumption that all particles get their velocity at the initial collision,
the Bjorken flow describes the boost-invariant longitudinal (\ie along the heavy-ion beams) 
expansion of the QGP. Because of its formulation, this flow is naturally described recurring to the 
Milne coordinate system, where the macroscopic velocity results at rest. 
The formulation of the flow in a static Minkowski space-time is on one hand more complex,
but potentially useful in the validation of codes working in a Cartesian laboratory frame coordinates.

In this work, we present the analytic solution for the Bjorken flow, 
considering a inviscid fluid consisting of both massive and massless particles,
working in a $(2+1)$ and $(3+1)$ Minkowski space-time coordinate system.

We provide details for the implementation of the benchmark using 
the Relativistic Lattice Boltzmann Method (RLBM)~\cite{gabbana-pr-2020},
a class of numerical models providing a computational efficient approach 
for the solution of the the relaxation time approximation of the relativistic Boltzmann equation.
Since lattice kinetic models rely on a mesoscopic description of the dynamics
viscous effects are naturally included, with relativistic invariance 
and causality preserved by construction (this at variance with respect to 
aforementioned hydrodynamic models).

Numerical results are shown to be in excellent agreement with the analytic solutions. 

This paper is organized as follows: in Sec.~\ref{sec:2-relativistic-hydrodynamics} 
we provide a brief introduction on the notation and the relevant equations 
for the kinetic description of a relativistic fluid.
In Sec.~\ref{sec:3-bjorken-flow} we provide analytic solutions for the Bjorken flow
in a Minkowski space-time for fluids of both massive and massless particles.
In Sec.~\ref{sec:4-results} we report the implementation of the benchmark 
with RLBM, and compare the numerical results with the analytic solutions
for a few selected cases.

Finally, conclusions are summarized in Sec.~\ref{sec:5-conclusions}.


\section{Ideal Relativistic Hydrodynamics}
\label{sec:2-relativistic-hydrodynamics}

\begin{figure}[tbh]
        \centering
        \includegraphics[width=0.8\textwidth]{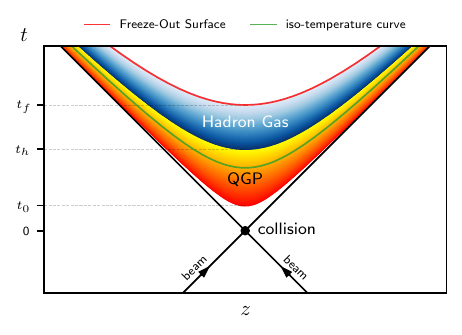}
        \caption{Schematic description of the phase transitions occurring in QGP in a
         z-t plane. At a time $t_0 \sim 1$ fm/c after the collision and at values of temperature 
         above the Hagedorn temperature, quarks become deconfined and free to roam in a so called gas
          of colored charges that interact with each other through the strong force. The resulting QGP 
        fireball expands and cools down until hadronization becomes possible again ($t_h \sim 5$ fm/c),
         and the QGP quickly turns into a gas of hadrons that undergoes further transformations as 
        neutrons and protons aggregate into nuclei (chemical freeze-out) and later on nuclei and 
        electrons aggregate into atoms (thermal freeze out, occurring at time $t = t_f$). This 
        point corresponds in the history of the Universe to the surface of last scattering, when 
        Cosmic Microwave Background was formed.}
        \label{fig:schema}
\end{figure}

In this section we consider a gas of particles with rest mass $m$ in a (d+1)-dimensional Minkowski 
space-time. The metric tensor is given by $\eta^{\alpha\beta} = \text{diag} (1, -\mathds{1})$,
with $\mathds{1} = \left( 1, \dots, 1 \right) \in \mathbb{N}^d $.

Einstein's summation convention is in place, with Greek indexes running from $0$ to $d$, and latin indexes 
running from $1$ to $d$ (spatial d-dimensional vectors are represented in bold). 

The starting point in the development of a relativistic kinetic theory is the single-particle 
distribution function $f(x^{\alpha}, p^{\alpha})$, where $x^\alpha = (ct, \bld{x})$ are space-time 
coordinates, and $p^\alpha = (p^0, \bld{p})$ is relativistic momentum, that accounts for the number 
of particles contained at time $t$ in the $2d$-dimensional phase space of infinitesimal volume 
$\diff \bld{x} \diff \bld{p}$.

An evolution equation for $f(x^{\alpha}, p^{\alpha})$ can be worked out considering the usual 
hypothesis of kinetic theory and the language of special relativity:

\begin{align}\label{eq:rbe}
        p^{\alpha} \frac{\dd f}{\dd x^\alpha} = -\frac{p^\alpha U_\alpha}{c^2 \tau} \big(f - f^\eq \big)    \; ,
\end{align}

the so called \textit{Relativistic Boltzmann Equation}, here written in the Anderson-Witting 
approximation~\cite{anderson-witting-ph-1974a}. $\tau$ is a typical relaxation time, \ie the 
typical time needed to $f$ to approach its equilibrium value $f^\eq$, the Maxwell-J{\"u}ttner 
~\cite{juettner-ap-1911} distribution

\begin{align}\label{eq:mjuettner}
        f^\eq = \left(\frac{c}{\kb T}\right)^d \frac{n}{2^{\frac{d+1}{2}} \pi ^{\frac{d-1}{2}} \zeta ^{\frac{d+1}{2}} K_{\frac{d+1}{2}}(\zeta )} exp{\left(-\frac{p^\alpha U_\alpha}{\kb T}\right)} \; ,
\end{align}

where $K_i(\zeta)$ is the modified Bessel function of the second kind of index $i$, $T(\bld{x},t)$ 
the temperature, $n(\bld{x},t)$ the particle number density, $U^\alpha(\bld{x},t)$ the macroscopic 
fluid velocity.

The parameter $\zeta = \frac{m c^2}{\kb T}$, named \textit{Relativistic Coldness}, is the ratio 
between the rest energy of a particle $m c^2$ and the thermal energy of the gas $\kb T$. This is the 
parameter that quantifies the level of 'relativity' of the gas:

\begin{align*}
        \zeta  \ll 1    \quad \Rightarrow &\text{ ultra-relativistic regime (massless particles)}  \\
        \zeta \sim 1    \quad \Rightarrow &\text{ mildly-relativistic regime}                       \\
        \zeta  \gg 1    \quad \Rightarrow &\text{ classical (non-relativistic) regime}
\end{align*}

The first and second order moments of the distribution function $f(x^{\alpha}, p^{\alpha})$ are the 
quantities of interest in relativistic hydrodynamics, the \textit{Particle Flow} $N^{\alpha}$ and the 
\textit{Energy-Momentum Tensor} $T^{\alpha\beta}$: 

\begin{align}
        N^\alpha        &= c \int p^\alpha         f \frac{\diff \bld{p}}{p^0} \label{eq:moments1} \; , \\
        T^{\alpha\beta} &= c \int p^\alpha p^\beta f \frac{\diff \bld{p}}{p^0} \label{eq:moments2} \; .
\end{align}

These quantities are conserved under collisions,

\begin{align}\label{eq:conservation-equations}
        \dd_\alpha N^\alpha = 0         \;, \quad\quad\quad     
        \dd_\alpha T^{\alpha\beta} = 0  \;,
\end{align}

and when the fluid is ideal they can be decomposed in the following way:

\begin{align}
        N^{\alpha}      &= n U^\alpha \label{eq:landau-decomp-N} \; , \\
        T^{\alpha\beta} &= (\epsilon -P) \frac{U^\alpha U^\beta}{c^2} + P \eta^{\alpha\beta}  
                                      \label{eq:landau-decomp-T} \; ,                 
\end{align}

where $P(\bld{x},t)$ is the hydrostatic pressure and $\epsilon(\bld{x},t)$ the energy density.

Lastly the following ideal Equation of State (EOS) is considered:

\begin{equation}\label{eq:eos}
\begin{array}{rll}
  \epsilon      &=& P \left( \zeta \frac{ K_{\frac{d+3}{2}}(\zeta) }{ K_{\frac{d+1}{2}}(\zeta) } - 1 \right) \; , \\ 
        P       &=& n \kb T                                                                                  \; .
\end{array}
\end{equation}     


\section{Bjorken Flow}\label{sec:3-bjorken-flow}

At sufficiently high energies ($\kb T \sim 1$ GeV - $1$ TeV), quarks and gluons become asymptotically 
free from the strong interaction which binds them into hadrons, and form a plasma of Partons which 
can be thought of as an extremely dense and hot gas of relativistic particles. 

The gas expands and cools down, until the hadronization temperature ($ \kb T \sim 170$ MeV, also 
called Hagedorn temperature~\cite{rafelski-book-2016}) is reached. Matter re-hadronize into baryons
and mesons. As the temperature further decreases, the gas goes through two additional phase 
transitions: at $\kb T \sim 100$ KeV protons and neutrons bind together to form atomic nuclei 
(\textit{Chemical Freeze-Out}), and at $\kb T \sim 1/4$ eV (\textit{Thermal Freeze-Out}) complete 
atoms are formed. 

The process here described, and sketched in Fig.~\ref{fig:schema}, is expected to provide a 
description of the first moments after the big bang~\cite{heinz-arxv-2004}. 

The Bjorken flow, also called the mono-dimensional boost invariant expansion model 
\cite{bjorken-prd-1983}, represents the simplest setup for modeling the longitudinal 
expansion of the dynamics observed in heavy ion collisions in particle accelerators.

It takes into consideration an inviscid fluid which is expanding along the 
propagation direction of the two beams (the z-axis cf.~Fig.~\ref{fig:schema}), 
with fluid velocity 
\begin{align}\label{eq:beta-t0}
  \beta = \frac{U^z}{U^0} = \frac{z}{c t}        \; .
\end{align}

In the next section, we detail the governing equations of the flow, and provide details on the 
analytical solutions. For simplicity, in what follow we make use of natural units: $c = \kb = 1$.

\subsection{Analytic Solution for an ideal fluid}
\label{subsec:3.1-bjorken-flow-solution}

We use a flat space-time, with coordinates $(t,z)$, where only the spatial coordinate
$z$ is considered since it is assumed that no transverse dynamic develops.

As previously mentioned, the most natural coordinate system for describing a Bjorken flow
is given by Milne coordinates, since in this metric the flow describes a
static fluid in a longitudinally expanding space-time.
Milne coordinates $(\uptau, w)$ are defined as
\begin{align}
        \begin{cases}
              \uptau &= \sqrt{t^2-z^2} \\
                   w &= \arctanh{\left(\frac{z}{t}\right)}
        \end{cases} \; .
\end{align}
In what follows, we denote with a tilde superscript quantities expressed in the Milne metric.

We define the following transformation matrix between the two reference frames:
\begin{equation}
  \Lambda^{\mu}_{\phantom{\mu}\nu} = \frac{\dd \tilde{x}^\mu}{\dd x^\nu} \quad ;
\end{equation}
the above provides an expression for the metric in the new basis:
\begin{align}
  \tilde{\eta}^{\alpha\beta} &= \Lambda_{\mu\phantom{\alpha}}^{\alpha}
                                                           \Lambda_{\nu\phantom{ \beta}}^{\beta} ~ \eta^{\mu\nu} 
                                                     = \text{diag} \left( +1, -\frac{1}{\uptau^2} \right) \; , \\
  \tilde{\eta}_{\alpha\beta} &= \text{diag} \left( +1, - \uptau^2 \right) \; .
\end{align}

Moreover, the curvilinear coordinates define Christoffel symbols $\Gamma^\alpha_{\beta\gamma}$, and 
the usual derivative is replaced with the covariant derivative: 
\begin{align}
  A^\alpha_{;\beta} = \partial_\beta A^\alpha + \Gamma^\alpha_{\beta\gamma}A^\gamma \; . 
\end{align}

Milne's Christoffel symbols are defined by 

\begin{align}
        \tilde{\Gamma}^{0}_{dd} = \uptau                  \quad\quad 
        \tilde{\Gamma}^{d}_{0d} = \frac{1}{\uptau}        \quad\quad                              
        \tilde{\Gamma}^{d}_{d0} = \frac{1}{\uptau}        \; ,
\end{align}
with all the other symbols being zero. 

From the condition in Eq.~\ref{eq:beta-t0}, we define the macroscopic velocity $U^\alpha$:
\begin{align}\label{eq:macro-vel}
  U^\alpha = \frac{1}{\sqrt{t^2-z^2}}\left( t, z \right)   \; .
\end{align}
By applying the transformation matrix $\Lambda^{\mu}_{\phantom{\mu}\nu}$ 
we obtain the corresponding Milne macroscopic velocity $\tilde{U}^\alpha$:
\begin{align}\label{eq:milne_velocity}
        \tilde{U}^\mu = \Lambda^{\mu}_{\phantom{\mu}\nu} U^\nu = (1, 0) \; .
\end{align}

The particle-flow (Eq.~\ref{eq:moments1}) and the energy-momentum tensor (Eq.~\ref{eq:moments2}) 
in Milne coordinates write as
\begin{align}\label{eq:milne_moments}
        \tilde{N}^{\alpha}      &= n \tilde{U}^{\alpha}  \; ,\\
        \tilde{T}^{\alpha\beta} &= 
        (\epsilon + P) \tilde{U}^\alpha \tilde{U}^\beta - P \tilde{\eta}^{\alpha\beta} \; .
\end{align}

Moreover, the balance equations (Eq.~\ref{eq:conservation-equations}) in Milne coordinates take 
the form

\begin{align}
        0 &= \tilde{N}^{\alpha}_{\phantom{\alpha};\alpha} 
           = \partial_{\alpha} \tilde{N}^{\alpha} + \tilde{\Gamma}^{\beta}_{\beta\alpha} \tilde{N}^{\alpha} \; , \\     
        0 &= \tilde{T}^{\alpha\beta}_{\phantom{\alpha\beta};\alpha}                                                             
           = \partial_{\alpha} \tilde{T}^{\alpha\beta} + \tilde{\Gamma}^{\beta}_{\mu\alpha} \tilde{T}^{\mu\alpha} 
           + \tilde{\Gamma}^{\alpha}_{\mu\alpha} \tilde{T}^{\mu\beta} \; .
\end{align}
 
By imposing such conservation equations, in combination with Eq.~\ref{eq:milne_velocity} and Eq.~
\ref{eq:milne_moments}, one gets:

\begin{align}\label{eq:bjorken_ode}
        0 &= \partial_\uptau (n \uptau)                                             \notag  \; , \\ 
        0 &= \partial_\uptau (\epsilon) + \frac{P+\epsilon}{\uptau}                         \; , \\
        0 &= \partial_w (P)                                                      \notag \;,
\end{align}

Combining the above with the EOS it is possible to 
derive analytic expressions for the particle number density $n$, 
the energy density $\epsilon$, and the temperature $T$.

Considering a ultra-relativistic gas, with ideal EOS 
\begin{align}
        P        &= n T                      \notag  \; , \\
        \epsilon &= d P     \label{eq:bjorken_eos_ultra} \; ,
\end{align}
the solution for the scalar fields $n, \epsilon, P, T$ can be obtained in the local 
Minkowski metric with a simple coordinate substitution, leading to:
\begin{align}\label{eq:sol_inviscid}
                n(\uptau)         &= n_0 \frac{\uptau_0}{\uptau} 
                                   = n_0 \left( \frac{t_0^2-z_0^2}{t^2-z^2} \right)^{1/2}               \notag  \; , \\
                \epsilon(\uptau)  &= \epsilon_0 \left(\frac{\uptau_0}{\uptau}\right)^{\frac{d+1}{d}}            
                                   = \epsilon_0 \left( \frac{t_0^2-z_0^2}{t^2-z^2} \right)^{\frac{d+1}{2d}}   \; , \\
                P(\uptau)         &= P_0 \left(\frac{\uptau_0}{\uptau}\right)^{\frac{d+1}{d}}           
                                   = P_0 \left( \frac{t_0^2-z_0^2}{t^2-z^2} \right)^{\frac{d+1}{2d}}    \notag  \; , \\ 
                T(\uptau)         &= T_0 \left(\frac{\uptau_0}{\uptau}\right)^{\frac{1}{d}}             
                                   = t_0 \left( \frac{t_0^2-z_0^2}{t^2-z^2} \right)^\frac{1}{2d}\notag  \; .
\end{align}
In the above, the values $(n_0,\epsilon_0,P_0,T_0)$ are prescribed at $\uptau=\uptau_0(t_0,z_0)$.

We remark that for the more general EOS in Eq.~\ref{eq:eos}, suitable for a fluid consisting of
massive particles $(\zeta \neq 0)$, it is necessary to perform numerical integrations of the 
equations in Eq.~\ref{eq:bjorken_ode}. Specifically, one takes the second of 
Eq.~\ref{eq:bjorken_ode} and expresses it using the EOS Eq.~\ref{eq:eos}:

\begin{align}
        \partial_\tau \left[ n T \left( \zeta(T) \frac{ K_{\frac{d+3}{2}}(\zeta(T)) }{ K_{\frac{d+1}{2}}(\zeta(T)) } - 1 \right) \right] 
        + 
        \frac{n T}{\tau} \left( \zeta(T) \frac{ K_{\frac{d+3}{2}}(\zeta(T)) }{ K_{\frac{d+1}{2}}(\zeta(T)) } \right) = 0 \; .
\end{align}
Combining the above with the equation for the density (Eq.~\ref{eq:bjorken_ode}) delivers
\begin{align*}
        - \frac{n T}{\tau} \left( \zeta(T) \frac{ K_{\frac{d+3}{2}}(\zeta(T)) }{ K_{\frac{d+1}{2}}(\zeta(T)) } -1 \right)
        + n
        \partial_\tau \left[ T \left( \zeta(T) \frac{ K_{\frac{d+3}{2}}(\zeta(T)) }{ K_{\frac{d+1}{2}}(\zeta(T)) } - 1 \right) 
        \right] + \frac{n T}{\tau} \left( \zeta(T) \frac{ K_{\frac{d+3}{2}}(\zeta(T)) }{ K_{\frac{d+1}{2}}(\zeta(T)) } \right) = 0 \; ,
\end{align*}
from which it directly follows,
\begin{align}\label{eq:sol_massive} 
        \partial_\uptau \left[ T \left( \zeta(T) \frac{ K_{\frac{d+3}{2}}(\zeta(T)) }{ K_{\frac{d+1}{2}}(\zeta(T)) } - 1 \right) \right] + \frac{T}{\uptau} = 0 \; . 
\end{align}


\section{Numerical Results}
\label{sec:4-results}

In this section we report numerical results for the Bjorken flow simulated in 
a $(2+1)$ and $(3+1)$ Minkowski coordinate system, using the Relativistic 
Lattice Boltzmann Method (RLBM).
We first give a short overview of the numerical method, 
followed by a comparison of the numerical results with the analytic
solutions presented in the previous section.

\subsection{Relativistic Lattice Boltzmann Method}
\label{subsec:4.1-rlbm}

RLBM is a computationally efficient approach to dissipative relativistic hydrodynamics.
The derivation of the method and its algorithm (see~\cite{gabbana-pr-2020} for full details), 
closely follow its non-relativistic counterpart~\cite{krueger-book-2016, succi-book-2018}:
RLBM solves a minimal version of Eq.~\ref{eq:rbe}, in which the 
discretization of the microscopic momentum vector on a Cartesian grid is coupled with
a Gauss-type quadrature  which ensures the preservation of the lower 
(hydrodynamics) moments of the particle distribution:
\begin{align}\label{eq:rlbm}
  f_i(\bm{x} + \bm{v}_i \Delta t, t+\Delta t) 
  =
  f_i(\bm{x}, t) - \Delta t\frac{p^\mu_i U^\mu}{c p_i^0 \tau} \left(f_i(\bm{x}, t) - f_i^{\rm eq}(\bm{x}, t) \right) \quad i = 1,2,\dots M \quad .
\end{align}
where $\bm{v}_{i} = \bm{p}_i / p^0_i$ are the (discrete) microscopic velocities, and 
$f_i^\eq{}$ is the discrete equilibrium distribution obtained from a polynomial expansion 
of Eq.~\ref{eq:mjuettner}:
\begin{align}\label{eq:feq_exp_trunc}
  f_i^{\rm eq}(\bm{x}, t) = w_i \sum_{k=1}^{N} a_{(k)}(U^\mu(\bm{x}, t),T(\bm{x}, t)) J^{(k)}(p^\mu_i) \quad ;
\end{align}
refer to Appendix~F and ~G in \cite{gabbana-pr-2020} for the definition of the 
polynomials and the projection coefficients used in the expansion.

Eq.~\ref{eq:rlbm} can be evolved in time following the collide-streaming 
paradigm typical of classic Lattice Boltzmann schemes. Moreover, at each time
step, and for each grid cell, we can compute the macroscopic fields from the moments
of the discrete particle distribution. 
Indeed, the quadrature rule allows the computation of integrals in Eq.~\ref{eq:moments1} and 
Eq.~\ref{eq:moments2} in a \textit{exact} form~\cite{gabbana-pr-2020, kovvali-book-2010}  via discrete summation:
\begin{align}\label{eq:discrete_sum_moments}
  N^\alpha(\bm{x}, t) = \sum_i p^\alpha_i f_i(\bm{x}, t)  \; , 
  \quad  T^{\alpha\beta}(\bm{x}, t) = \sum_i p^\alpha_i p^\beta_i f_i(\bm{x}, t) \; .
\end{align}

Finally, all the macroscopic fields can be determined combining the above 
with the eigenvalue problem 
\begin{equation}
  T^{\alpha \beta} U_{\beta} = \epsilon U^{\alpha} \; ,
\end{equation}
which allows calculating the density $n$ via
\begin{equation}
  n = \frac{1}{c^2} N^{\alpha} U_{\alpha} ,
\end{equation}
and finally the temperature $T$ from the EOS in Eq.~\ref{eq:eos}.

\subsection{Simulation Details and Results}
\label{subsec:4.1-rlbm}

We consider a domain of lenght $L = 1~\rm{fm}$ along the $z$ dimension, with 
$z \in [-\frac{L}{2},\frac{L}{2}]$, discretized using $2000$ grid points, and with one single grid 
point used to represent the other (periodic) dimensions.

We follow the dynamics from an initial time value of $t_0 = 1~\rm{fm/c}$, up to $t_h = 5~\rm{fm/c}$, 
which approximately represents a time in the QGP dynamics when a departure from a hydrodynamic 
regime occurs~\cite{paquet-npa-2017}.

Initial prescriptions for the temperature and the density are given at time $t_0$ at the value 
$z_0 = L/2 \leq t_0$, which in turn defines 

\begin{align}\label{eq:tau_initialization}
       \uptau_0 = \sqrt{t_0^2 - z_0^2} \; .
\end{align}
\begin{figure}[tbh]
        \centering
        \includegraphics[width=0.93\textwidth]{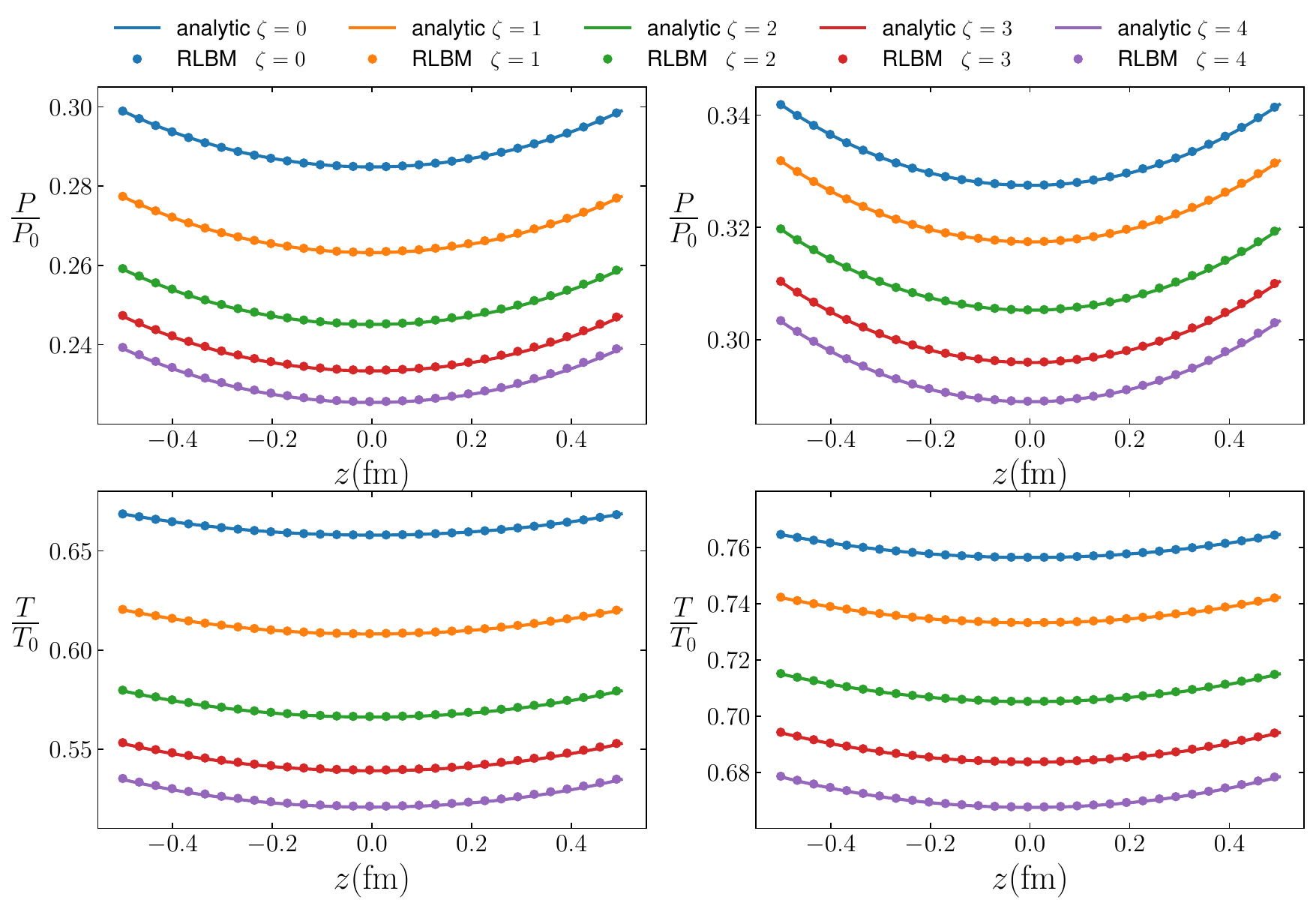}
        \caption{ Comparison of numerical results for the Bjorken flow of an 
        ideal gas of particles with different values of the relativistic coldness $\zeta$, 
        computed with respect to the reference temperature $\kb T_0 = 300$ MeV. Represented in the 
        figure are the Pressure and Temperature profiles in the $z$ spatial domain, taken at time 
        $t = 2$ fm/c, both considering a $d=2$ (left column) and $d=3$ (right column) ideal Equation
         of State. The numerical results are confronted with the analytic ($\zeta = 0$) and 
         semi-analytic ($\zeta \neq 0$) results, and a perfect match is obtained.}
        \label{fig:bjorken_results_cartesian}
\end{figure}
We take as initial value for the temperature $T(t_0, z_0) = T_0 = 300~\rm{MeV}$, that is 
approximately the value given by both theoretical~\cite{bjorken-prd-1983} and lattice QCD 
~\cite{fodor-jhp-2004} calculations, and is well above the Hagedorn temperature 
~\cite{rafelski-book-2016}. 
Following~\cite{ambrus-prc-2018}, we set the particle number to $n(t_0, z_0) = n_0 = 1.5~\rm{fm}^{-d} $.

We select the following values for the relativistic coldness $\zeta = m c^2 / \kb T_0$:   
$\zeta = (0,1,2,3,4)$, that with the value chosen for $T_0$ roughly translate to rest masses 
that fall within the quark mass range~\cite{particle-data-group-jpg-2010}. 

The macroscopic fields $n(t_0,z)$, $T(t_0,z)$ are initialized from Eq.~\ref{eq:sol_inviscid} 
(or by numerically solving Eq.~\ref{eq:sol_massive} in the case $\zeta \neq 0$),
while the macroscopic velocity $U^\alpha(t_0,z)$ is set according to Eq.~\ref{eq:macro-vel}. 

The discrete particle distribution functions $f_i$ are initialized at equilibrium
\begin{align} 
        f_i(t_0,z)=f^{\eq}_i \left( n(t_0,z), T(t_0,z), U^\alpha(t_0,z) \right) \quad .
\end{align}

We implement boundary conditions along the $z$ axis using the following expression:
\begin{equation}
  f_i\left(t, z=\pm \left(\frac{L}{2}+dz \right)\right) 
  = 
  f^{\eq}_i \left( n_*(t,z), T_*(t,z), U_{*}^\alpha(t,z) \right) + \phi_i \quad ,
\end{equation}
where quantities denoted with the subscript ``*'' are calculated from Eq.~\ref{eq:sol_inviscid} 
and Eq.~\ref{eq:macro-vel},
and with $\phi_i$ a zero-th order extrapolation of the non-equilibrium part of the distribution
calculated from the inner grid points.
\begin{figure}[tbh]
        \centering
        \includegraphics[width=0.93\textwidth]{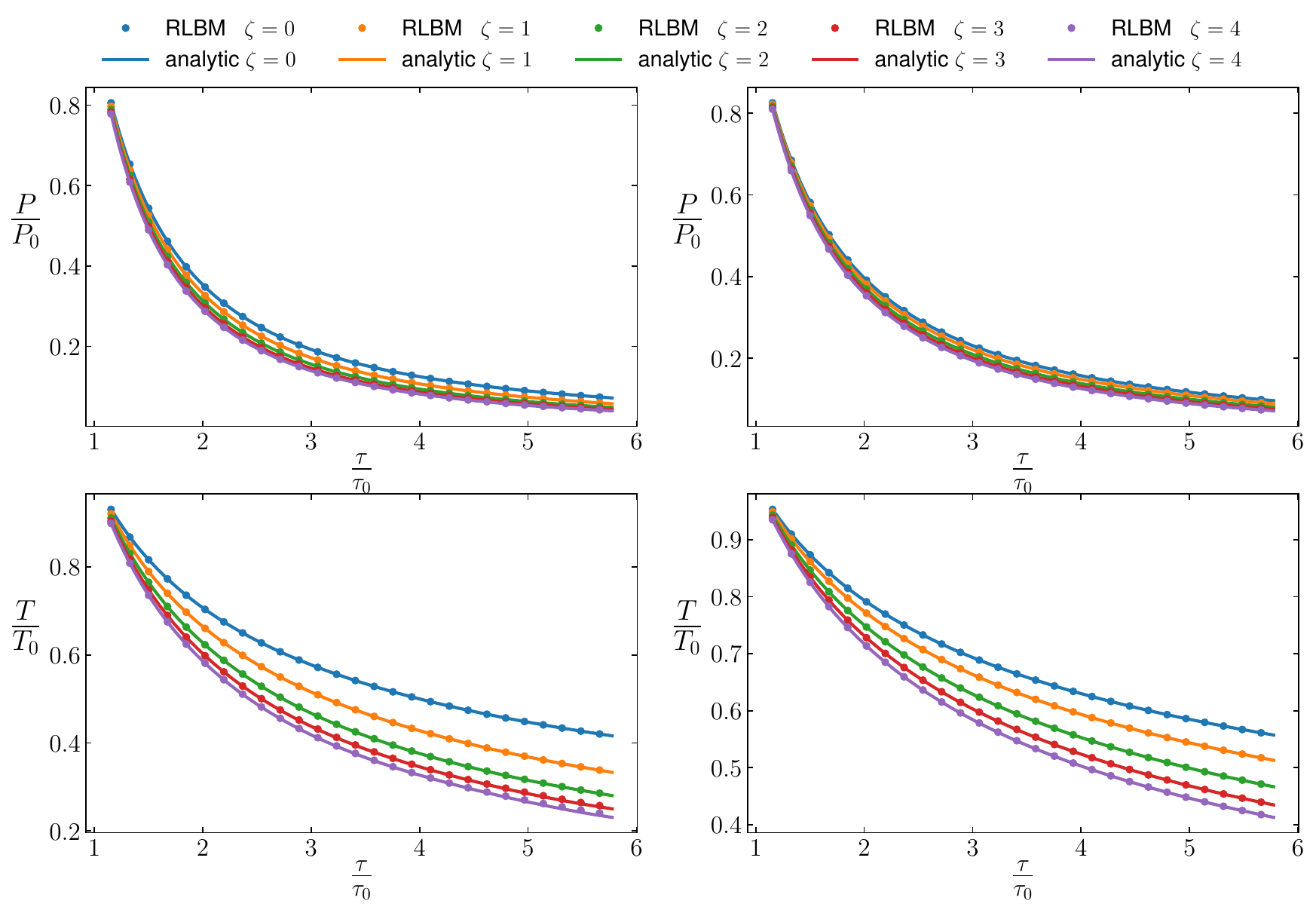}
        \caption{ Comparison of numerical results for the Bjorken flow of an 
        ideal gas of particles with different values of the relativistic coldness $\zeta$, 
        computed with respect to the reference temperature $\kb T_0 = 300$ MeV. Represented in the 
        figure are the Pressure and Temperature profiles in the $\uptau$ time domain, both considering
         a $d=2$ (left column) and $d=3$ (right column) ideal Equation of State. The initial value 
         for $\uptau_0$ is computed according to \ref{eq:tau_initialization}, where $L=1$ fm and 
         $t_0 = 1$ fm/c.
        The numerical results are confronted with the analytic ($\zeta = 0$) and semi-analytic 
         ($\zeta \neq 0$) results, and a perfect match is obtained.}
        \label{fig:bjorken_results_milne}
\end{figure} 

In Fig.~\ref{fig:bjorken_results_cartesian} and Fig.~\ref{fig:bjorken_results_milne} we compare 
numerical results from RLBM simulations against the analytic solutions presented in 
Sec.~\ref{sec:3-bjorken-flow}.

In Fig.~\ref{fig:bjorken_results_cartesian} we show the profiles for temperature and pressure
as a function of the spatial coordinate $z$. The snapshots are taken at $t = 2~\rm{fm/c}$.
The macroscopic profiles are in excellent agreement with the analytic solution. 

In the panels on the left hand side we report results in $(2+1)$ dimensions, while on the right 
we show the results in $(3+1)$ dimensions. Although the flow is mono-dimensional, the 
number of spatial dimensions enters the definition of the EOS~\ref{eq:eos}.
We observe that the evolution of the QGP phase is accelerated in the $(2+1)$-dimensional case,
with the gas cooling down at a quicker rate with respect to the $(3+1)$ counterpart.

Moreover, from Fig.~\ref{fig:bjorken_results_cartesian} it is possible to appreciate that the 
evolution of fluids consisting of heavier particles leads to lower temperature values and therefore 
to a quicker cooling of the QGP. This suggests that such cases would lead to phase transitions at 
an earlier time, in complete consistence with theoretical calculations and experimental observations
~\cite{waqas-sr-2021} and providing a basis for more complex coalescence models 
~\cite{fries-annurev-2008}.

Finally, in Fig.~\ref{fig:bjorken_results_milne} we give a different representation of
Fig.~\ref{fig:bjorken_results_cartesian}, with the macroscopic profiles presented
as function of the Milne coordinate $\uptau$. We once again stress that the RLBM used in this 
work operates in Minkowski coordinates; the results in Fig.~\ref{fig:bjorken_results_milne}
have been translated into the Milne spacetime via a simple coordinate transformation
thanks to the fact that all the thermodynamic fields are Lorentz scalars.


\section{Conclusions}
\label{sec:5-conclusions}

In this work, we have presented analytic and numerical solutions describing a Bjorken 
flow in a flat space-time coordinate system. This type of benchmark provides a simplified
description of the longitudinal expansion of the QGP, mimicking the dynamics
observed in relativistic heavy ion collisions taking place in particle colliders such as 
RHIC and LHC.

We have considered the dynamics of a relativistic fluid for several kinematic parameters,
in particular varying the rest mass of the particles, as well as the equation of state.
The numerical results show excellent agreement with the analytic solutions.

This work may provide a useful reference for evaluating the accuracy of numerical solvers
for relativistic hydrodynamics working in a Minkowski flat space-time. Furthermore, the 
proposed solver might be useful for further simulations of QGP, that go beyond the 
longitudinal description and investigate the transversal motion as well. 
As an example, study of elliptic flows~\cite{snellings-njp-2011} and characterizations 
of Bjorken attractors~\cite{blaizot-2021-plb,ambrus-prd-2021} might be natural next steps of 
investigation.


\section*{Acknowledgments}
\label{sec:6-Acknowledgments}

SS acknowledges funding from the European Research Council under the European Union’s
Horizon 2020 framework programme (No. P/2014-2020)/ERC Grant Agreement No. 739964 (COPMAT). 
We would like to thank Victor Ambru\ifmmode \mbox{\c{s}}\else \c{s}\fi{} for helpful discussions.
All numerical work has been performed on the COKA computing cluster at Università di Ferrara.


\bibliographystyle{elsarticle-num}
\bibliography{biblio}

\end{document}